\documentclass[twocolumn]{aa}
\usepackage{graphicx,psfig,epsfig,epsf}
\usepackage{amssymb}
\usepackage{txfonts}
\usepackage{color}
\definecolor{red}{rgb}{0.7,0,0}
\definecolor{blue}{rgb}{0,0,0.7}

\begin{document}
\def\igr{IGR~J19140$+$0951 }

   \title{Unveiling the Nature of the High Energy  Source IGR~J19140$+$0951}

   \author{J. Rodriguez\inst{1,2},  C. Cabanac\inst{3},  D.C. Hannikainen\inst{4}, V. Beckmann \inst{5,6}, S.E. Shaw\inst{7,2}, J. Schultz\inst{4}}

   \offprints{J. Rodriguez:jrodriguez@cea.fr}

   \institute{Centre d'Etudes de Saclay, DAPNIA/Service
              d'Astrophysique (CNRS FRE 2591), Bat. 709, Orme des
              Merisiers, Gif-sur-Yvette Cedex 91191, France 
         \and
             {\it INTEGRAL} Science Data Centre, Chemin d'\'Ecogia 16,
              CH-1290 Versoix, Switzerland 
          \and
            Laboratoire d'Astrophysique, Observatoire de Grenoble, BP
              53X, 38041 Grenoble, France
 	    \and
             Observatory, PO Box 14, FIN-00014 University of
              Helsinki, Finland 
	      \and
	     NASA Goddard Space Flight Center, Code 661, Building 2 
Greenbelt, MD 20771, USA
            \and Joint Center for Astrophysics, Department of Physics, University of Maryland, Baltimore County, MD 21250, USA
            \and School of Physics and Astronomy, University of Southampton, Southampton, SO17 1BJ, UK
             }

\authorrunning{J. Rodriguez et al.}
\titlerunning{On the Nature of IGR~J19140$+$0951}
   \date{}

   \abstract{We report on high energy observations of IGR~J19140$+$0951 performed with {\it RXTE} on three
occasions in 2002, 2003 and 2004, and {\it INTEGRAL} during a very well sampled and unprecedented high energy 
coverage of this source from early-March to mid-May 2003.  Our analysis shows that \igr spends most of its 
time in a very low luminosity state, probably corresponding to the
state observed with {\it RXTE}, and 
characterised by thermal Comptonisation. In some occasions we observe variations of the luminosity by a 
factor of about 10 during which the spectrum can show evidence for a thermal component, besides 
thermal Comptonisation by a hotter plasma than during the low luminosity state. The spectral parameters 
obtained from the spectral fits to the {\it INTEGRAL} and {\it RXTE} data strongly suggest that \igr 
hosts a neutron star rather than a black hole. Very importantly, we observe variations of the absorption 
column density (with a value as high as  $\sim 10^{23}$ cm$^{-2}$). 
Our spectral analysis also reveals  a bright iron line  detected with 
both {\it RXTE}/PCA and {\it INTEGRAL}/JEM-X, at different levels of luminosity. We discuss these 
results and the behaviour of IGR~J19140$+$0951, and show, by comparison with other well known systems 
(Vela X-1, GX 301$-$2, 4U 2206+54), that \igr is most probably a High Mass X-ray Binary.

}

   \maketitle
%

\section{Introduction}

\indent \igr was serendipitously discovered during the first INTErnational Gamma-Ray Astrophysical
Laboratory ({\it INTEGRAL}, Winkler et al. 2003) observation of the Galactic microquasar 
GRS~1915+105 (Hannikainen, Rodriguez \& Pottschmidt 2003). Inspection of the high energy 
archives showed it to be the most likely hard X-ray counterpart of the poorly
studied  {\it EXOSAT} source EXO~1912+097 (Lu et al. 1996). 
Soon after its discovery a Target of Opportunity (ToO) was performed on \igr with 
the {\it Rossi X-ray Timing Explorer (RXTE)}. The preliminary spectral analysis
of this ToO showed the source had a rather hard spectrum, fitted 
with a power law of photon index 1.6, and an absorption
column density N$_{\mathrm{H}}$=6$\times10^{22}$~cm$^{-2}$ (Swank \& Markwardt
2003). Recently timing analysis of the {\it RXTE}/ASM data revealed an 
X-ray period of 13.55 days (Corbet, Hannikainen \& Remillard 2004). This analysis
showed that the source was detected even during the early days of the {\it RXTE} mission, which 
suggests that \igr is a persistent X-ray source  although most of the time in a faint state. 
In a companion paper (Hannikainen et al. 2004a, hereafter Paper 1)
we used the latest version of the {\it INTEGRAL} software to refine
and give the most accurate X-ray position of \igr (see also Cabanac et al. 2004),
which allowed us to obtain the most accurate X-ray/Gamma-ray spectra of the 
source. High energy spectral analysis of \igr covering the period of
its discovery, {\it i.e.}. during  {\it INTEGRAL} revolution
48,  was  presented for the first time.  
We have, in particular, shown, that during this observation, the source,
although very variable, showed two distinct spectral behaviours.
The first one manifests a  thermal component (black body-like) in the soft
X-rays, and a  hard X-ray tail, whereas the second one is harder
and can be interpreted as originating from thermal Comptonisation (Paper 1).\\
\indent Although it is very likely that \igr is a  Galactic object, the nature
of the compact object is unclear. The spectral analysis presented in Paper 1 would tend 
to favour a neutron star, but no definite conclusion could be drawn from the
data presented.\\
\indent We report here observations of \igr with {\it INTEGRAL} performed
between March and May 2003, during a very well sampled and unprecedented  high energy 
coverage of this source. To our {\it INTEGRAL} monitoring, we  
add the analysis of the March 2003 {\it RXTE} ToO, as well as the analysis of observations 
performed one year earlier on EXO 1912+097, and  the first {\it RXTE} observation of a monitoring
campaign we are currently leading on IGR~J19140+0951, performed in April 2004. 
We  describe the 
sequence of observations and the data reduction procedures in Sec. 2, and then 
present the results obtained from the different instruments in Sec. 3. The results are  discussed
 in the last part of the paper.


\section{Observations and data reduction}
\begin{figure}[htbp]
\epsfig{file=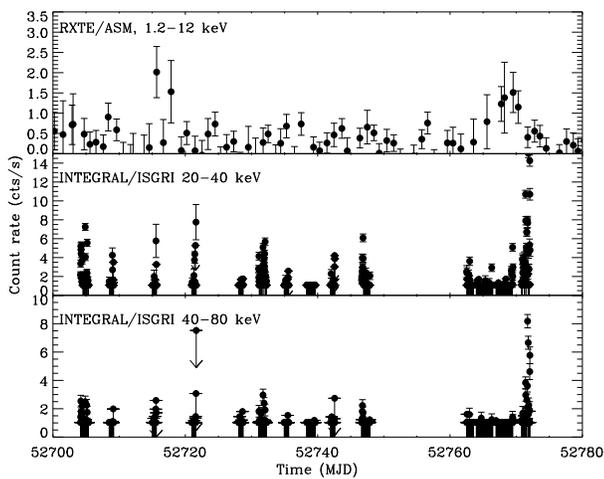,width=\columnwidth}
\caption{{\bf top}: One-day average 1.2--12 keV {\textit{ASM}} light curve. We show 
here only the data covering the 2003 March--May period.
{\bf middle \& bottom}: 20--40 keV and 40--80 keV {\textit{ISGRI}} light curves over the same period. Note that 
 the Crab luminosity corresponds to 75 cts/s for the ASM, and it is $\sim 114$ cts/s and
$\sim 67$ cts/s for ISGRI, in the 20--40 and 40--80 keV ranges respectively.}
\label{fig:lc}
\end{figure}

The 1.2--12 keV {\textit{RXTE/}}ASM, 20--40 keV and 40--80 keV {\textit{INTEGRAL}}/ISGRI
light curves of the source covering the period of interest are shown in Fig.
\ref{fig:lc}.
\subsection{{\it INTEGRAL}}
\label{sec:integraldata}
The main instruments on board {\it INTEGRAL} are IBIS (Ubertini et
   al. 2003), and SPI (Vedrenne et al.  2003). Both
   instruments use coded masks which allow $\gamma$-ray imaging over
   large Fields of View (FOV), $\sim30 \times30^\circ$
   up to zero response.
  The Totally Coded FOV (TCFOV) is a smaller part within which the
   detector has the highest response.  IBIS has a TCFOV of $9^\circ\times9^\circ$, while that of 
 SPI is 16$^\circ$ (corner to corner).
The {\it INTEGRAL} Soft Gamma-Ray Imager (ISGRI, Lebrun et al. 2003) is the
  top layer of the IBIS detection plane, 
  and covers the energy range from 13~keV to a few hundred keV.\\
\indent The JEM-X monitors (Lund et al. 2003), consist 
  of two identical coded mask instruments designed for X-ray imaging in
  the range 3--35~keV with an angular resolution of 3 arcmin and a
  timing accuracy of 122~$\mu$s. The JEM-X FOV is smaller with a diameter of the fully-coded 
FOV of  $4.8^\circ$. 
  During our observation only the JEM-X unit 2 was being used.\\
\indent We focus here on the monitoring of the source performed between its (re-)discovery by
{\it INTEGRAL}, in March 2003 (Hannikainen et al. 2003) and May 2003, during 
which we obtained an unprecedented high energy coverage of the source.
The journal of the {\it INTEGRAL} observations is presented in Table~\ref{tab:log}.
Among the data acquired by our team with GRS~1915+105 as the main target (e.g. 
Hannikainen et al. 2004b), we obtained data through exchange with several other 
teams. Therefore, while in the former group of observations \igr is always 
in the totally coded FOV of IBIS (and thus in the FOV of JEM-X), in the 
latter group of data the source lies in any position of IBIS, and 
is most of the time outside the JEM-X FOV. Note that 
in addition to those guest observer data, \igr was observed during {\it INTEGRAL} 
Science Working Team ToOs on GRS~1915+105 (Fuchs et al. 2003). Those data are also 
included in this study. It should be noted that an {\textit{INTEGRAL}} observation
consists of a sequence of pointings (or science windows, hereafter SCW) following
a certain pattern around the main target of the observation on the plane of the
sky (Courvoisier et al. 2003). The patterns are also reported in Table \ref{tab:log}. 
The chosen  pattern has some importance on the amount of useful data. 
For an on-axis source (or close to as is \igr in our revolutions), the 
hexagonal pattern allowed us to have the source always in the JEM-X FOV, while 
this is not so for  the $5\times5$ pattern. When the source is far 
off-axis, it may even be outside the FOV of IBIS in some SCW.
  
\begin{table}[htpb]
\caption{Journal of the {\it INTEGRAL} observations presented in the paper. 
$^\dagger$(Mean) Angle between \igr and the centre of the field of view. The 
total exposure represents the time spent by {\it INTEGRAL} on each
field. 
}
\begin{tabular}{cccccc}
\hline
\hline
Rev. \# & Start & Stop & Observing & Total Exposure & off-axis  \\
        & (MJD) &  (MJD) & Pattern &         & angle$^\dagger$    \\
\hline
48 & 52704.12 & 52705.37 & Hexagonal & 101 ks & 1.1$^\circ$ \\
49 & 52708.45 & 52709.16 & 5$\times$5& 55 ks & 9.3$^\circ$ \\
51 & 52715.00 & 52715.66 & 5$\times$5& 55 ks & 9.3$^\circ$ \\
53 & 52720.95 & 52721.62 & 5$\times$5& 55 ks & 9.3$^\circ$ \\
56 & 52728.04 & 52728.79 & 5$\times$5& 55 ks & 9.3$^\circ$ \\
57 & 52731 04 & 52732.29 & 5$\times$5& 101 ks & 1.1$^\circ$ \\
58 & 52734.91 & 52735.62 & 5$\times$5& 55 ks & 9.3$^\circ$ \\
59 & 52738.29 & 52739.41 & Hexagonal & 101 ks & 1.1$^\circ$ \\
60 & 52741.91 & 52742.62 & 5$\times$5& 55 ks & 9.3$^\circ$ \\
62 & 52746 62 & 52747.91 & 5$\times$5& 101 ks & 1.1$^\circ$ \\
67 & 52762.50 & 52763.54 & 5$\times$5& 84 ks & 4.9$^\circ$ \\
68 & 52763.95 & 52766.45 & 5$\times$5& 200 ks & 4.9$^\circ$ \\
69\_1 & 52766.91 & 52768.37 &5$\times$5& 117 ks & 4.9$^\circ$\\
69\_2 & 52768.41 & 52769.5 & Hexagonal & 88 ks & 1.1$^\circ$\\
70 & 52770.79 & 52772.08 & 5$\times$5& 100 ks & 4.9$^\circ$ \\
\hline
\hline
\end{tabular}
\label{tab:log}
\end{table}

\begin{table}[htbp]
\centering
\caption{Level of systematic uncertainty applied to the spectral 
channels of the JEM-X spectra, and energy channel correspondence.}
\begin{tabular}{ccc}
\hline
\hline
Channel & Energy & Systematic uncertainty\\
        & (keV)  &   (\%)\\
\hline
58--79 & 4.00--5.76        & 5 \\
80--89 &  5.76--6.56       & 2 \\
90--99 &  6.56--7.36       & 7 \\
100--109 & 7.36--8.16      & 5 \\
110--119 & 8.16--9.12      & 4\\
120--129 & 9.12--10.24      & 5\\
130--149 & 10.24--13.44      & 4\\
150--159 & 13.44--15.40      & 6\\
160--169 & 15.40--17.64      & 5\\
170--179 & 17.64--20.24      & 8\\
180--189 & 20.24--22.84      & 7\\
190--197 & 22.84--25.52      & 9\\
\hline
\end{tabular}
\label{tab:sysjemx}
\end{table}

\indent The JEM-X data were reduced using the Off-line Scientific
Analysis (OSA) 4.1 software, following 
the standard procedure described in the JEM-X cookbook. 
Due to the faintness of IGR~J19140$+$0951 we forced the source extraction
at the position  reported in Cabanac et al. (2004). 
We ran the analysis on all the revolutions considered here 
when IGR~J19140$+$0951 was in the JEM-X FOV
(48, 57, 59, 62, 67, 68, 69, 70) but only included the SCWs where the 
source was at an offset angle less than 5$^{\circ}$. The level 
of systematic uncertainty applied to each spectral channel, and the energy-channel
correspondence is reported in Table \ref{tab:sysjemx}
 (P. Kretschmar \& S. Mart\'{\i}nez N\'u\~nez priv. comm.). 
\indent The IBIS/ISGRI data were reduced using version 4.1 of  
  the OSA software. The data reduction procedure is identical to the one described
in Paper 1, i.e. for each revolution we first ran the software up to the production 
of images and mosaics in the 20--40 and 40--80 keV energy ranges. The software was here 
free to find the most  significant sources in the images. We then created a catalogue containing
only the 9 brightest sources of the field (either detected in some of the revolutions or 
in all), and re-ran the software forcing the extraction of the count-rate of those sources.
The data products obtained through the ISGRI pipeline therefore include 20--40 keV and
40--80 keV light curves (Fig. \ref{fig:lc}), with a time bin about 2200s 
(typical length of a SCW). Rather than using the standard spectral extraction, we
extracted spectra from images/mosaics accumulated at different times. This
non-standard method and its validity is described in Appendix 1.\\
\indent First of all we restricted the spectral analysis to the times when
the source was both in the IBIS and JEM-X FOV, {\it i.e.} revolutions 48, 57, 59, 62, 
67, 68, 69 and 70.  The distinction of the different times was defined from the 20--40 keV 
light curve (Fig. \ref{fig:lc}), on a SCW basis in a way similar to what is presented 
in Paper 1. The distinction of different times to accumulate the data from is solely based 
on the level of luminosity of the source during a SCW. Although the level on which the distinction
is made is rather arbitrary, our approach allows us to try to understand the origin of the 
variability on the time scale of a SCW by accumulating spectra of similar (hard) luminosity. Although this
  approach can hide and completely miss the spectral variations on smaller time scales, it is 
  dictated by the need to accumulate a large number of JEM-X and IBIS spectra to obtain good constraints 
  on the spectral parameters (e.g. Paper 1). Our PCA analysis (Sec. 3.2) shows that although the source
 can  be variable on short time scales, the fitting of the average spectrum leads to a rather good
 representation of the physics underlying the source emission.   Here
due to a larger sampling of the source as compared  to Paper 1, it was
possible to define more ``states'' to accumulate the spectra from, 
in order  to understand better the origin of the variations and try to avoid possible mixture of different
 states together. We define here:
\begin{itemize}
\item The ``ultra faint'' state was accumulated from all SCW when the source had a 
20--40 keV count rate (CR, measured in cts/s) $<$ 1. 
\item The ``faint state'' has a similar definition as in Paper 1 and was 
accumulated from all SCW where $1\leq \mathrm{CR} < 3$.
\item The ``bright state'' corresponds to  $3\leq \mathrm{CR} < 6$.
\item The ``ultra bright state'' corresponds to the bright 20--40 keV flares, {\it i.e.}
 $\mathrm{CR} \geq 6$.
\end{itemize}
We caution the reader that these definition of ``states'' have nothing to do with the standard 
definition of spectral states usually employed in studies of X-ray 
Binaries (e.g. Tanaka \& Shibazaki 1996), and that they refer to luminosity in the hard X-rays. 
We thus extracted the source count rate and error from  20 bin mosaics accumulated during 
these four intervals as described in Appendix 1. 6\% systematics have been 
added to all spectral channels.
The JEM-X individual spectra were averaged together following the same time distinction.\\
\indent We also tentatively extracted SPI spectra following the standard method.
However, the SPI angular resolution is about 2$^\circ$, which renders the analysis 
of \igr delicate given the proximity to GRS~1915+105, which is much brighter 
(Hannikainen et al. 2004b, Rodriguez et al. 2004a). In fact, an analysis of the SPI spectra 
showed that the parameters were consistent with those of GRS~1915+105.
We therefore did not include the SPI data in our analysis.\\
\indent The JEM-X \& ISGRI spectra were then fitted in XSPEC v11.3.1, with latest
rmf file for JEM-X ({\tt jmx2\_rmf\_grp\_0021.fits}), and the OSA 3.0 ISGRI matrices for 
 IBIS ({\tt isgr\_rmf\_grp\_0010.fits, isgr\_arf\_rsp\_0004.fits}). We retained the energy 
channels between 4 and 25 keV for JEM-X and those between 20 and 150 keV 
for ISGRI. Further rebinning of the JEM-X data was applied so that both ISGRI and 
JEM-X data give similar weight to the $\chi^2$ statistics in the spectral fittings.

\subsection{RXTE data}
The field of \igr has been observed 3 times with  {\it RXTE} during
pointed observations,  the journal of which is summarised in Table \ref{tab:log2}. 
Two observations were truly dedicated to IGR~J19140+0951, a public ToO, 
and an observation that is part of an on-going monitoring programme of the source. 
The third and oldest observation  was dedicated to EXO 1912+097. Whether or not \igr and the 
{\it EXOSAT} source are the same is beyond the scope of this paper, given 
that the best position of \igr (Cabanac et al. 2004) is still consistent with the 
{\it EXOSAT} position of EXO 1912+097 (Lu et al. 1996).  
We  assume in the following that the sources are the same.\\
\indent The {\it RXTE} data have been reduced with the {\it LHEASOFT} package v5.3.1,
following the standard procedures for both {\it Proportional Counter Array} 
(PCA, Jahoda et al. 1996), and {\it High Energy Timing Experiment} 
(HEXTE, Rothschild et al. 1998) data. See e.g. Rodriguez et al. (2003a, 2004b)
for the procedure of spectral extraction, and 2--40~keV (channel 5--92) high time resolution 
light curves. In addition, and since the source is quite weak, we further 
rejected times of high electron background in the PCA (i.e. times when the electron ratio in Proportional 
Counter Unit (PCU) $\#2$ is greater than 0.1), and time during the passage through the South Atlantic 
Anomaly (i.e. we retained the times since SAA$> 30$ or $< 0$ minutes) to 
define the ``good time intervals'', and used the latest background files
available for faint sources. The spectra were extracted from the top layer of all 
PCUs turned on during each observation. In order to account for uncertainties in the 
response matrix we added 0.8\% systematics below 8 keV, and 0.4\% above (Rodriguez et al. 2003a).  
Note that during the three observations, the data formats were different resulting in 
different time resolutions for the timing study. We could explore the source temporal 
behaviour up to  64~Hz, 4000~Hz, and 124~Hz in Obs. 1, 2, and 3 respectively.
  For HEXTE, we separated on and off source pointings 
and carefully checked for any background measurement pointing on GRS~1915+105, 
and other close-by sources (XTE J1908+094, X 1908+075, \& 4U 1909+07).
We only used  the pointings which were 
not contaminated by other  sources as background maps. However, due to either the weakness
of \igr or the limited number of background maps, no HEXTE data can sensibly be used in our analysis.
 We therefore focus on a comparison of the PCA spectra obtained during the 3 observations
The  spectra were fitted  in XSPEC V11.3.1 (Arnaud 1996), between 3 and 
25 keV.
\begin{table}
\caption{Journal of the {\it RXTE} observations discussed in the paper. $^\dagger$ Net 3-25 keV mean count rate (cts/s)/PCU, only the top layers of PCUs are considered.}
\begin{tabular}{ccccc}
\hline
\hline
Obs. \# & MJD  & Exposure & \# PCU & Count rate/PCU$^\dagger$\\
\hline
1 & 52394.08  & 3248 s & 2 & 6.8 \\
2 & 52708.79  & 2848 s & 4 & 6.6 \\
3 & 53087.50  & 6496 s & 3 & 11.8\\
\hline
\hline
\end{tabular}
\label{tab:log2}
\end{table}

\section{Results}
\subsection{High resolution temporal analysis}
We studied the PCA high resolution light curves in different frequency
ranges given the different time resolution of the different data format, in order to investigate 
the time variability and search for Quasi-Periodic Oscillations. 
We produce 2--40 keV Power Density Spectra (PDS)  on an interval length of 16~s. Our PDS were 
normalised according to Leahy et al. (1983). The lower boundary of the PDS is 
0.0625~Hz in each case while the higher boundary is 64~Hz, 4000~Hz, and  128~Hz for 
Obs. 1, 2 and 3 respectively.
The 3 PDSs are well fitted with constants with best values   
1.999$\pm 0.007$ ($\chi^2_\nu=1.10$ for 117 DOF),  2.002$\pm0.001$ ($\chi^2_\nu=1.09$ 
for 199 DOF) and 2.004$\pm0.004$ ($\chi^2_\nu=0.84$ for 139 DOF) 
(error at the 90\% confidence level), compatible with the expected 
value for purely Poisson noise. In case a  High Frequency QPO (HFQPO) 
is present it usually has a higher rms amplitude 
at energy higher than $\sim5$--$7$~keV. We also produced 
a PDS in the 7--20 keV range from Obs. 2,
and analysed it between 0.0625~Hz and 4000~Hz. A single constant fits the PDS well, with best 
value 2.000$\pm0.001$ (at 90\% confidence level), again indicative of purely Poisson noise.\\
Using \begin{equation}
\mathrm{R}=\sqrt{2\times\mathrm{n_{\sigma}}\times\frac{\mathrm{S+B}}{\mathrm{S}^2}
\sqrt{\frac{\Delta\nu}{\mathrm{T}}}} 
\label{eq:limit}
\end{equation} where R is the fractional rms amplitude, S is the source 
net rate, B is the background rate, T and $\Delta\nu$ are the exposure time and the width
 of the QPO,  one can estimate the 3$\sigma$ upper limit on the detection
of any QPO at any frequency, during the 3 observations. The limiting amplitude 
being proportional to the square root of the width, the limit for a sharp QPO will be 
lower than that of a broad feature. The most constraining results are obtained for Obs. 3, 
for which the limit on the presence of a Q(=$\nu/\Delta\nu$)=10 low frequency 
feature is comprised between $1.5\%$ ($\nu=0.0625$~Hz) and  6.5\% ($\nu=20.0$~Hz). 
This puts tight constraints on the presence of such a feature since those low frequency QPOs are 
usually observed to have a rather high fractional amplitude (e.g. 5--30\%, McClintock \& Remillard 2004). 
For high frequency QPOs,  however, the situation is reversed. With the help of Equation \ref{eq:limit},
 we obtain a limit of 17.4 \% for a 200~Hz QPO during Obs. 2. This means that if such a feature was 
 present (in the 100--300 Hz range for a black hole and in the kHz range for a neutron star) then 
 we would miss it. This is even true for a $\sim 15\%$ rms HFQPO as
 sometimes detected in some Atoll sources (Swank 2004). 

\subsection{Spectral Analysis}
\begin{figure*}[htbp]
\epsfig{file=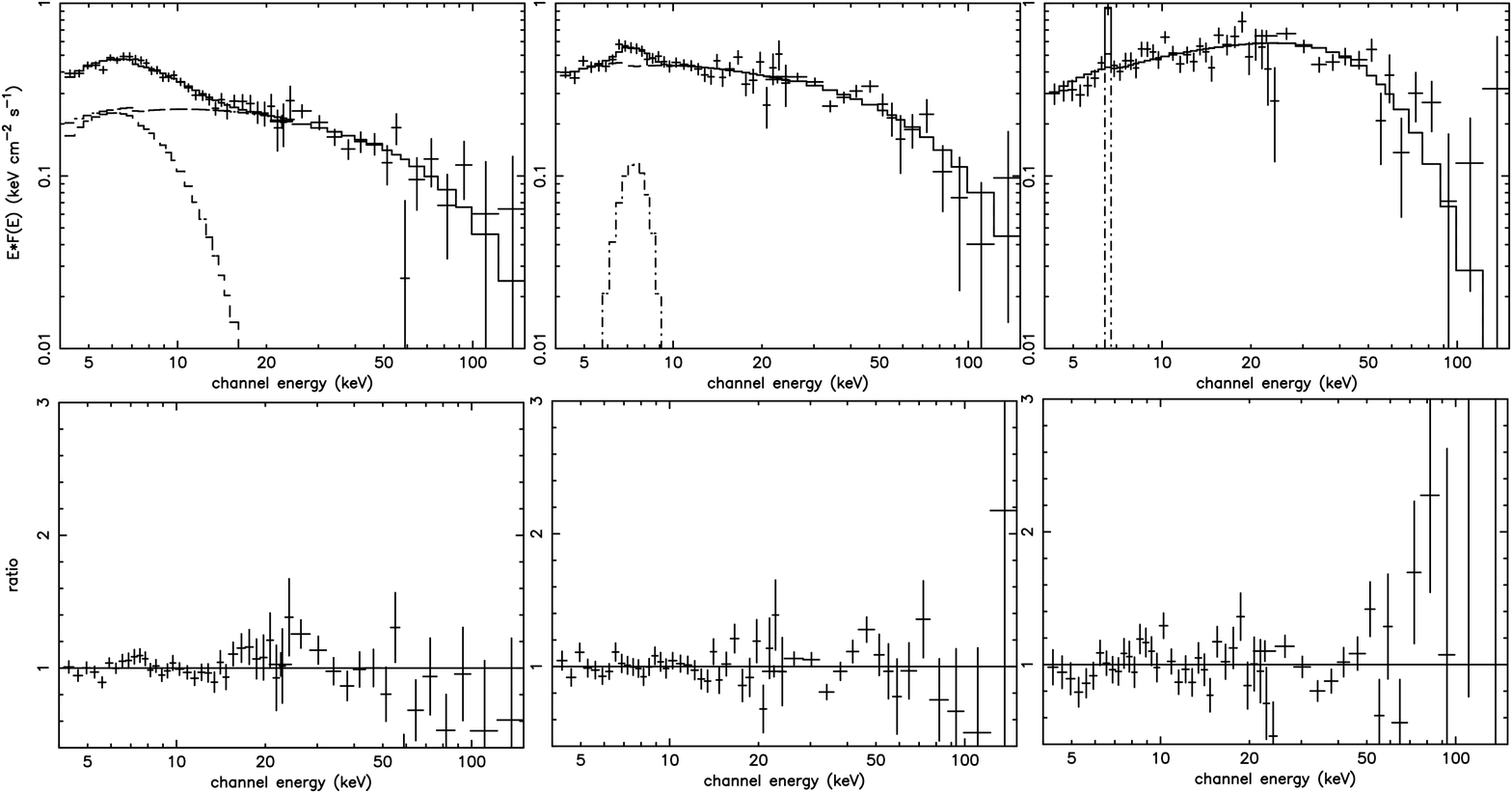,width=18cm}
\caption{{\it{INTEGRAL}}/JEM-X and ISGRI spectra with the best (physical) model superimposed in each case (see 
the text for details). The different component of the models ({\tt comptt}, black body or Gaussian) are also plotted.
From left to right, the panel correspond to the ``Faint state'', the ``Bright state'', the ``Ultra bright state''.
The vertical axis is in $\nu-F_\nu$ units. The lower panels represent the ratio between the model and
the data in each of the "states".}
\label{fig:integspec}
\end{figure*}
\begin{figure*}
\centering
\epsfig{file=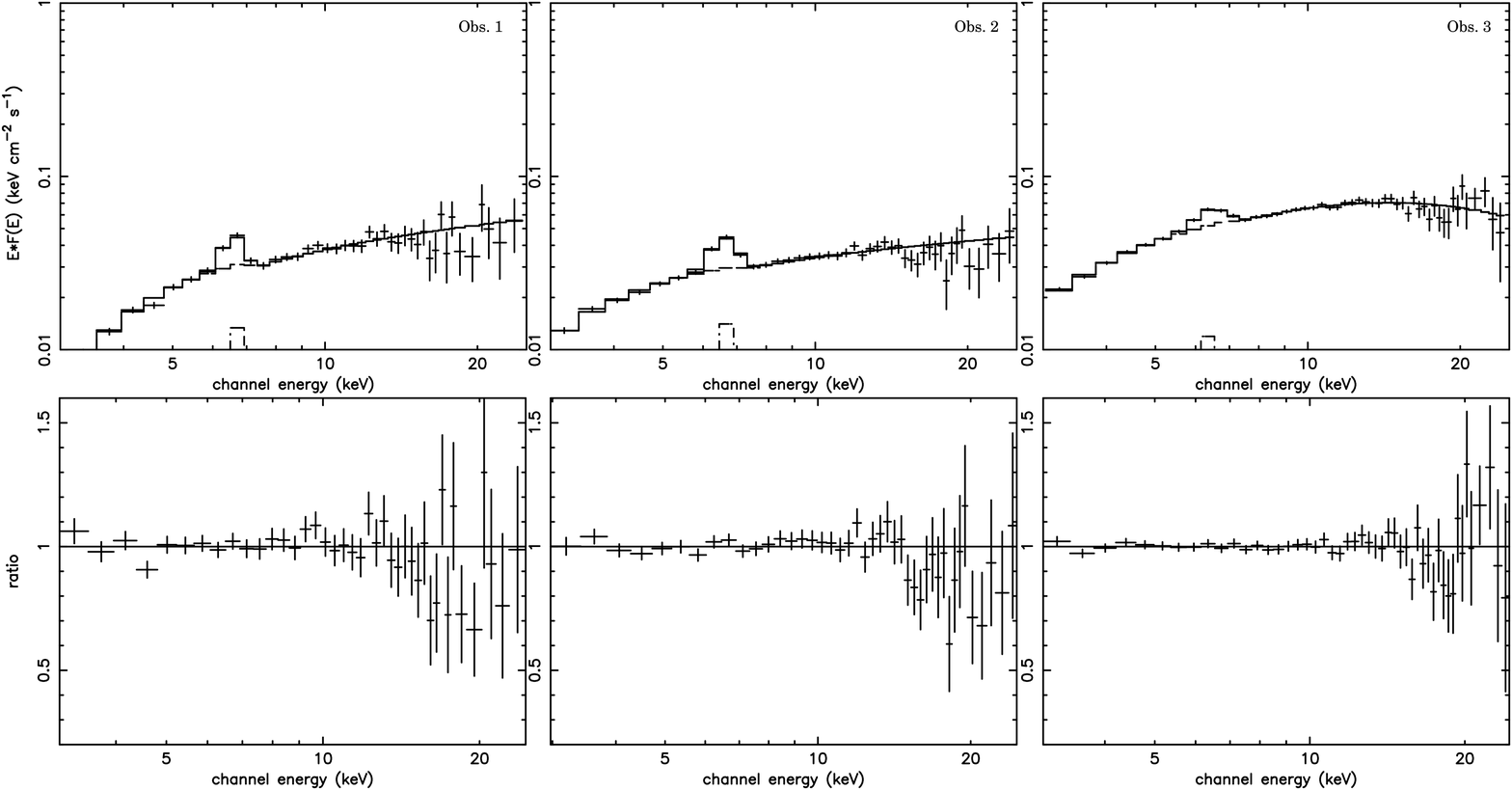,width=18cm}
\caption{{\it RXTE}/PCA spectra of EXO 1912+097/\igr with the best fit model superimposed: 
a Gaussian at $\sim6.5$ keV,
plus an absorbed power law for the left and middle one, and an absorbed power law with a high energy cut-off 
for the right one. The spectra are time ordered from left to right (1 year separation). The Gaussian 
is represented in all 3 spectra. Note that the same vertical scale as that of Fig. \ref{fig:integspec} is employed
to facilitate the comparison. The lower panels represent the ratio between the best 
model and the data.}
\label{fig:pcaspec}
\end{figure*}
\subsubsection{Simultaneous JEM-X/ISGRI spectral analysis}
\label{sec:integspec}
Over a total of 450 SCW, covering revolutions 48, 57, 59, 62, 67, 68, 69 and 70, 
\igr is found in the ``ultra faint'' state during 271 SCW (60.2\%),
it is in the ``faint state'' during 130 SCW ($\sim 28.9\%$), in the ``bright state'' 
during 37 SCW (8.2\%), and in the ``ultra bright'' state during 12 SCW ($\sim2.7\%$).
However, due to the 5$\times$5 observing pattern (Courvoisier et al. 2003) and mean off 
axis angle during revolutions 67, 68, beginning of 69, and 70 (Table \ref{tab:log}), 
the source is outside of the JEM-X FOV, during a large part of these revolutions. 
For the sake of consistency, we extracted mean spectra from the time when \igr is in 
both the ISGRI and JEM-X FOV. However, in doing so some statistical sensitivity is lost 
especially at high energies, and we completely miss the flare occurring at the end of 
revolution 70 (Fig. \ref{fig:lc}). Finally, the selection based on the availability of 
JEM-X (good) data leads to effective exposures of  $\sim 15$~ks, $\sim 55$~ks, 
$\sim 186$~ks, and $\sim 400$~ks for the ``ultra bright'', ``bright'', ``faint'', and ``ultra faint''  
states respectively.
In all our spectral fits a constant was included to take into account the cross 
calibration uncertainties, and was found at a similar value.\\
\indent Following the procedure presented in Paper 1, we first fitted the spectra from both 
instruments simultaneously, with a simple model consisting of an absorbed power law. 
The value of ${\mathrm{N}}_{\mathrm{H}}$, was frozen to the value obtained with RXTE 
(Swank \& Markwardt 2003), i.e. $6\times10^{22}$~cm$^{-2}$, since the useful
energy range of JEM-X does not allow us to obtain a better constraint on 
this parameter. We note, however, that this parameter may change from one observation 
to the other (see our {\it RXTE} spectral analysis below), but the results of our 4--150
keV spectral analysis remain largely unchanged, this energy range being largely unaffected
by absorption.\\
Since significant evolution at least in terms of luminosity, and possibly in terms of 
spectral parameters (Paper 1) is expected, we present here the results of the spectral fits
to the different ``states'' separately. \\
\indent {\bf{``Ultra Faint State'':}} The source is not detected in any of the spectral channels
of our ISGRI mosaic. Therefore it is not possible to construct a spectrum.
 We therefore did not include these data in our analysis since consistent comparison with the 
other states was not possible (mainly due to the lack of constraints on the possible hard tail,
cut-off etc.)

\indent {\bf{``Faint State'':}} The simple power law model gives a poor fit to the data with 
 a reduced chi square 
$\chi^2_\nu=2.56$ (47 DOF). Following the results from Paper 1, we added a black body 
component to the power law. This component is required at more than $5\sigma$.  
The best fit parameters are reported in Table \ref{tab:faintres}.
Replacing the power law by a cut-off power law ({\tt cutoffpl} hereafter 
CPL) slightly improves the fit (the cut-off is required at just the 
$3\sigma$ level), but the cut-off energy  is poorly constrained 
(E$_{\mathrm{cut}}=44_{-18}^{+44}$~keV) (all along the text errors are given at the
 90\% confidence level). A good fit
 is also achieved with a simple power law and a Gaussian ($\chi^2_\nu=1.26$ for 
44 DOF). The photon index is compatible ($\Gamma=2.32_{-0.08}^{+0.06}$) with the value 
obtained with the former model. The line parameters are those reported in Table 
\ref{tab:line}. Note that the large uncertainty on the line parameters, its large width and 
normalisation could indicate a possible mixing of line and the black body emission, as will 
be discussed in Section \ref{sec:line}. This possibility could explain well the inability 
of our fits to converge to sensible results when trying to fit the data including both the 
black body and the Gaussian.\\
\indent We tentatively replaced the phenomenological models with more physical models of 
Comptonisation. Using the {\tt comptt} model (Titarchuk 1994) alone does not provide
a good fit to the data. As in the previous case, adding a black body
 component 
improves the fit significantly. The temperature of the seed photon for Comptonisation tends to too
low values to be constrained. It is therefore frozen to 0.3 keV. The black body temperature is 
consistent with that obtained with the phenomenological model (kT=$1.42\pm0.06$~keV). Note that if a Gaussian
instead of the black body is added to the {\tt comptt}, a good fit can be achieved, but the
parameters of the line are not physically acceptable (the centroid
 tends to too low a value, while
the width is too high). The JEM-X and ISGRI spectra are plotted with the {\tt comptt+bbody} model  
superimposed in Fig.~\ref{fig:integspec}, left panel. \\

\begin{table*}[htbp]
\centering
\caption{Best fit parameters obtained  for the 
``Faint state'' observed with {\it INTEGRAL}. Errors are given at the 90\% level. Fluxes are in units of 
erg~cm$^{-2}$~s$^{-1}$}
\begin{tabular}{cccccc}
\hline
\hline
{\tt bbody+po}     & kT or kT$_\mathrm{e}$ &  $\Gamma$ or $\tau$& $\chi^2_\nu$ & \multicolumn{2}{c}{Unabs. flux}\\
             & (keV) &      & (DOF) & 1-20 keV & 20-200 keV\\

             & 1.51$\pm0.07$ & $2.35_{-0.08}^{+0.15}$ &1.36 (45) &$2.24 \times 10^{-9}$&
	     $5.19\times 10^{-10}$ \\      
\hline
{\tt bbody+comptt} &  21$_{-8}^{+1}$ & 1.3$_{-1.0}^{+0.2}$ & 1.07 (44) & $1.83\times 10^{-9}$ &
$4.26 \times 10^{-10}$ \\
\hline
\hline
\end{tabular}
\label{tab:faintres}
\end{table*}

\indent {\bf{``Bright State'':}} Here again the simple model of an absorbed power law does 
not fit the data well ($\chi^2_\nu=2.78$ for 47 DOF). A cut-off component is not required at 
a high level ($\gtrsim 3\sigma$). A black body and a simple power law 
does not provide  a good fit to the data. In fact, 
 an alternative model of a power law with  high energy cut-off and a Gaussian line 
provides a good fit to the data. The addition of the Gaussian
  leads to an improvement $\Delta\chi^2=36$ for $\Delta$DOF=3.  
  The best fit parameters for this state are reported in Table \ref{tab:brightres}, 
while the line parameters are discussed in Section \ref{sec:line}.
Note that besides the presence of the line, the spectral parameters are consistent 
with those presented in Paper 1.

\begin{table*}[htbp]
\centering 
\caption{Best fit parameters obtained with  the different spectral model for the ``Bright'', 
and ``Ultra Bright'' states observed with {\it INTEGRAL}. Errors are 
given at the 90\% level. Fluxes are in units of erg~cm$^{-2}$~s$^{-1}$. Note that 
a Gaussian line is included in the fit in all models. CPL stands for {\tt cutoffpl} 
in XSPEC terminology}
\begin{tabular}{cccccccc}
\hline
\hline
State & model&   $\Gamma$ &  E$_{\mathrm{cut}}$ or kT$_{e}$ &  $\tau$ &  $\chi^2_\nu$& \multicolumn{2}{c}{Unabs. flux}\\
       &      &         & (keV)  & & (DOF) & 1-20 keV & 20-200 keV\\
\hline
Bright & CPL            & $2.05_{-0.14}^{+0.08}$ & 71$_{-17}^{+29}$ & & 1.48 (43)& $2.52 \times
10^{-9}$& $7.32\times 10^{-10}$ \\  
       &{\tt comptt} &      & 22.0$_{-5.0}^{+15.6}$ & $1.2_{-0.6}^{+0.5}$  &  1.44 (43) & $2.42 \times 10^{-9}$ & 
       $7.12\times 10^{-10}$ \\  
\hline
Ultra Bright & CPL & $1.37_{-0.07}^{+0.14}$ & $27.1_{-4.7}^{+6.7}$ &  & 1.49 (43) & $1.93
\times 10^{-9}$& $9.54\times 10^{-10}$ \\  
             & {\tt comptt}       & & 11.2$_{-0.5}^{+0.8}$ & $2.97_{-0.15}^{+0.36}$ &  
	     1.75 (43) & $2.09 \times 10^{-9}$& $1.03\times 10^{-9}$ \\
\hline
\hline
\end{tabular}
\label{tab:brightres}
\end{table*}
Fitting the data with the {\tt comptt} alone leads to $\chi^2_\nu=2.04$ for 45 DOF. Again a  
black body 
component is marginally detected ($\gtrsim 3\sigma$). As with the phenomenological model, 
the fit is  greatly improved if  a Gaussian instead of the black body is added to the 
{\tt comptt} model. The Gaussian parameters are compatible with those found with the 
phenomenological model. Note that the temperature of the seed photon for 
Comptonisation is too low to be well constrained. We therefore fixed it at 0.3~keV.  
The best fit parameters are reported in Table \ref{tab:brightres}, while the spectra 
are shown in Fig. \ref{fig:integspec} middle panel.
 
\indent {\bf{``Ultra Bright State'':}} As in the other ``states'' the single component model 
does not represent the data well ($\chi^2_\nu=3.61$ for 47 DOF). A high energy cut-off is 
required at more than $5\sigma$. Adding a black body does not bring significant improvement. 
On the other hand,  adding a Gaussian improves 
the fit slightly ($\Delta\chi^2=15$ for $\Delta$DOF=3). The best fit parameters are reported 
in Table \ref{tab:brightres}, while the line parameters are discussed in Section 
\ref{sec:line}. Note that alternative models involving black body emission (either with 
a Gaussian and/or a high energy cut-off) do not provide a good description of the data. 
As in the ``Bright'' state the {\tt comptt} provides an acceptable fit if a Gaussian 
is added to the model. The temperature of the seed photons for Comptonisation
is again fixed at 0.3~keV. The line 
parameters are consistent with those found with the phenomenological model. 
The best fit parameters are reported in Table \ref{tab:brightres}, 
while the broad band spectrum is shown in Fig. \ref{fig:integspec} right panel.

\subsubsection{PCA spectral analysis} 
During the 3 {\it RXTE} observations the source was dimmer than when detected 
with {\it INTEGRAL} (see e.g. the differences between Fig. \ref{fig:integspec} and \ref{fig:pcaspec}). 
We fitted the spectra with the same spectral models, first a simple absorbed power law,
 or simple absorbed black body or disc black body. While the latter models give a 
poor description of the data, the former (after addition of a Gaussian at $\sim 6.4$~keV
to account for an excess due to Fe K complex emission) represents the data well 
for Obs. 1 and 2. The addition of the Gaussian leads to
  $\Delta\chi^2=56$, $\Delta\chi^2=101$ for $\Delta$DOF=3 in Obs. 1 and
  2 respectively.  The
best fit parameters are reported in Table \ref{tab:rxtefit}. The equivalent 
absorption column density ($N_{\mathrm H}$) was let free to vary in all spectral fits, and 
we note a slight decrease of $N_{\mathrm H}$ from Obs. 1 to Obs. 2 , the latter being consistent with 
the results reported by Swank \& Markwardt (2003).\\
\indent The simple power law+Gaussian model fails to represent Obs.  3 ($\chi^2_\nu=3.0$, 44 DOF). 
 Replacing the power law by a CPL leads to  a good fit, an F-test indicates the cut-off is required
   at more than 5$\sigma$. Note that the CPL alone does not provide
   a good fit to the data ($\chi^2_\nu=2.85$ for 46 DOF). The value of $N_{\mathrm H}$ is slightly lower than during Obs.  2 
(Table \ref{tab:rxtefit}). Because a cut-off power law is usually interpreted 
as a signature of thermal Comptonisation, we replaced the phenomenological model by
the {\tt comptt} model. This more physical model represents the data well,
but we note, however, that due to the 3 keV lower boundary of the PCA spectra, the 
input photon temperature is very poorly constrained ($<1.12$ keV at 90\%, if it is left as 
a free parameter). We then froze this parameter to 0.3 keV in a second run. The best 
parameters are reported in Table \ref{tab:rxtefit}. We note here that the value of 
$N_{\mathrm H}$ is more consistent with that obtained during Obs. 2.
 The line parameters are discussed in Section \ref{sec:line}.\\
\indent We then re-performed the fits to Obs.  1 and 2, either adding a high energy cut-off 
(with {\tt{highecut}}) or replacing the power law by a CPL. The 
improvement to the fits is only marginal (just at the $3\sigma$ level)
 for Obs. 1, and  $\gtrsim 3\sigma$ for Obs.  2, therefore not at high significance. 
We also replaced the phenomenological models by {\tt comptt}, and although a good fit 
is achieved the parameters (especially the 
electron temperatures) are found to be quite high and very poorly constrained.
The three spectra and the best fit models (simple power law for the first two and CPL
 for Obs. 3) are plotted in Fig.~\ref{fig:pcaspec}, together 
with the ratio between the model and the data.

\begin{figure}[htbp]
\epsfig{file=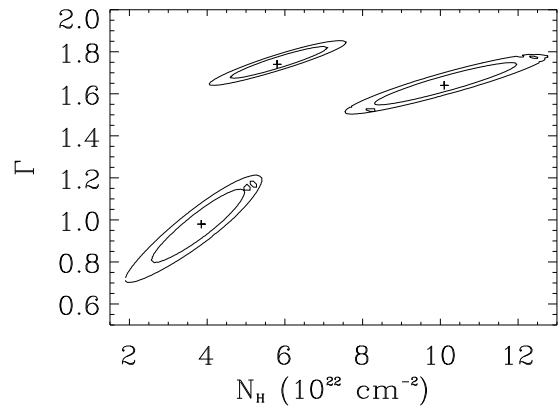,width=\columnwidth}
\caption{Error contours for the column density ($N_{\mathrm H}$) and the power law 
photon index ($\Gamma$) derived from the fits to the three {\it RXTE}/PCA 
spectra. The crosses mark the location of the best fit values, and the 68\% and 90\% confidence contours
are shown.}
\label{fig:contour}
\end{figure}

Since the model parameters (especially the power law photon index) are strongly correlated to 
the value of  $N_{\mathrm H}$, we represent the error contours of the photon index $\Gamma$ vs. 
the value of  $N_{\mathrm H}$, for the three observations in Fig.~\ref{fig:contour}.\\
\indent In addition to the simple power law fit, and in order to compare with the results
from the fits to the {\it INTEGRAL} data, we tentatively fitted the spectra with the {\tt comptt} model.
 The best fit parameters are reported in Table \ref{tab:rxtefit}.

\begin{table*}[htbp]
\caption{Best fit parameters obtained from the spectral fits to the {\it RXTE}/PCA data. The errors are
reported at the 90\% confidence level.
PL stands for power law, CPL for {\tt cutoffpl} in XSPEC terminology.}
\begin{tabular}{cccccccc}
\hline
\hline
Obs. Number & model & $N_{\mathrm H}$ & $\Gamma$ & $\mathrm{E_{cut}}$ or kT$_{e}$ &  $\tau$ & unabs. 1-20 keV flux & $\chi^2_\nu$ \\
            &       &  $\times 10^{22}$~cm$^{-2}$ &          &  (keV) / (keV) & (keV) / & erg~s$^{-1}$~cm$^{-2}$ & (DOF)\\  
\hline
1 & PL & 10.1$\pm 2.0$ & 1.64$\pm 0.11$ & & &$1.60\times10^{-10}$ & 0.95 (42)\\
\hline
2 & PL & 5.8$_{-1.3}^{+1.4}$ & 1.74$\pm0.08$ & & & $1.47\times10^{-10}$& 0.99 (42)\\
\hline
3 & CPL & 3.85$_{-1.5}^{+1.2}$ & 0.98$_{-0.21}^{+0.17}$ & 13.7$_{-2.6}^{+3.9}$ & & $2.37\times10^{-10}$ & 0.95 (40)\\
  & {\tt comptt}& 5.8$_{-1.0}^{+0.9}$ &  & 5.0$_{-0.4}^{+0.5}$ & 5.3$_{-0.4}^{+0.5}$ &$2.42\times10^{-10}$ & 0.99 (43)\\
\hline
\hline
\end{tabular}
\label{tab:rxtefit}
\end{table*}

\subsubsection{A closer look at RXTE observation 3}

\begin{figure*}[htpb]
\centering
\epsfig{file=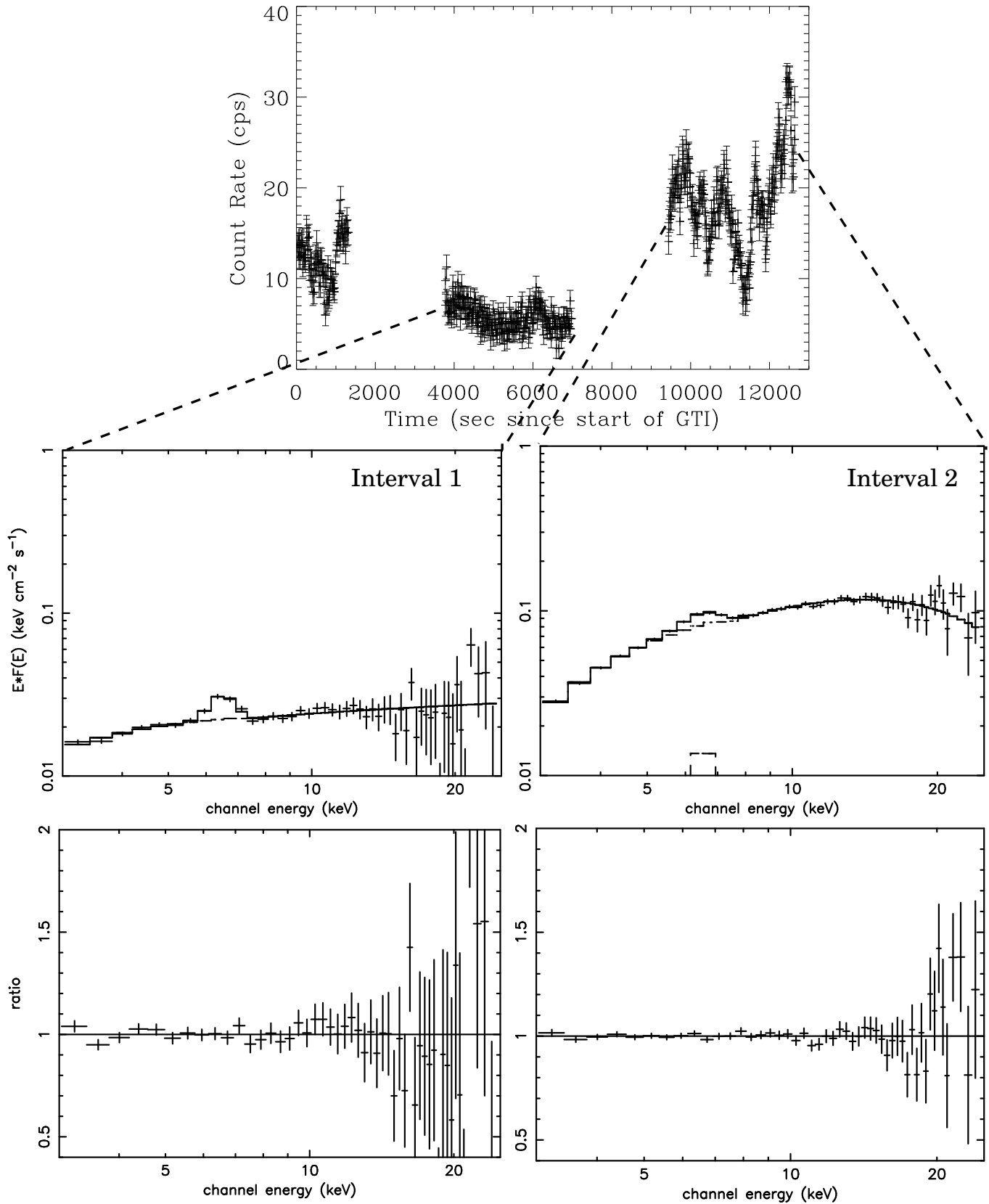,width=18cm}
\caption{{\bf{TOP}}:  2-40 keV PCA (Top Layer of PCU 2) light curve 
of Obs.  3.  {\bf{MIDDLE \& BOTTOM}}: {\it Left:} PCA 3-25 keV 
spectrum of interval 1, and its best fit model, the ratio between the model and
the data is represented below. {\it Right:} Same as left but for interval 2. 
Note that the same vertical scale is employed in both to facilitate the comparison.}
\label{fig:PCAHXT}
\end{figure*}

 Since the 16~s PCA light curve shows that the source
is very variable on short time scales, we  separated the observation into two periods,
one corresponding to the low and steady flux (second interval in Fig. \ref{fig:PCAHXT} left), 
and the other one corresponding to the high flux and large variations 
(third interval in Fig. \ref{fig:PCAHXT} left).\\
\indent  We applied the same (simple) 
models as discussed in the previous section. While for the first interval a simple 
absorbed power law (plus a Gaussian) fits the data well, the same model yields a 
poor fit for the second ($\chi^2_\nu$= 3.48 (44 DOF)). A cut-off improves the fit,
 and is required at  more than 5$\sigma$. The best fit results are reported 
in Table \ref{tab:RXTE}, while the spectra and best fit models are shown in 
Fig. \ref{fig:PCAHXT}. In this case again the {\tt comptt} fits the data well. 
The temperature of the seed photons is again fixed at 0.3 keV. As for the global spectrum, we 
remark that the absorption column returned from the fit with this model is slightly higher than 
the value obtained with CPL.  

\begin{table*}[htbp]
\centering
\caption{Best fit parameters to the PCA fits of the two intervals from Obs. 3.}
\begin{tabular}{cccccccc}
\hline
\hline
Interval & model & $N_{\mathrm H}$ & $\Gamma$ & $\mathrm{E_{cut}}$ or $kT_{\mathrm e}$& $ \tau$ & unabs. 1-20 keV flux & $\chi^2_\nu$ \\
            &       &  $\times 10^{22}$~cm$^{-2}$ &         &  (keV) &  & erg~s$^{-1}$~cm$^{-2}$ & (DOF)\\  
\hline
1 & PL & 2.5$_{-1.3}^{+1.0}$& 1.86$_{-0.10}^{+0.06}$& & &1.09$\times10^{-10}$  & 0.87 (44)\\
2 & CPL & 5.4$_{-1.2}^{+1.0}$& 0.86$_{-0.18}^{+0.15}$& 11.9$_{-1.8}^{+2.3}$ &  & 3.45$\times 10^{-10}$
&  0.83 (43)\\
  & {\tt comptt} & 7.0$_{-1.2}^{+1.0}$ & & 4.7$\pm0.3$ & 5.8$_{-0.4}^{+0.6}$ & 3.84$\times 10^{-10}$ &
  0.92 (43)\\
\hline
\end{tabular}
\label{tab:RXTE}
\end{table*}

We note a significant evolution of the absorption column density and of the power law photon
index between the two intervals. In order to check whether the evolution of both was real, 
we re-performed the fits freezing $N_{\mathrm H}$ to its mean value (Table \ref{tab:rxtefit}).
The spectral parameters obtained for both fits are compatible with those found leaving all 
parameters free to vary, except the power law photon index which tends to a softer value in 
interval 1 ($\Gamma=2.11\pm 0.05$), and a to harder one for interval 2 ($\Gamma=1.38\pm0.03$). 
Since $N_{\mathrm H}$ and $\Gamma$ are tightly correlated, we also re-performed the fit
freezing $\Gamma$ to its mean value, and allowing $N_{\mathrm H}$ to vary. While for interval 2 
the spectral parameters obtained in this case are close to the ones obtained when everything is 
free to vary, this method yields a poor fit for interval 1. We take these results as 
evidence that both $\Gamma$ and  $N_{\mathrm H}$ vary between both intervals. Note 
that this likely variation of the absorption is reinforced by the variations of $N_{\mathrm H}$
we observe between Obs. 1, 2 and 3 (Table \ref{tab:rxtefit}).
 
\subsubsection{The iron line}
\label{sec:line}
As mentioned previously in all the {\it INTEGRAL} and {\it RXTE} spectra, an 
iron K$\alpha$ fluorescence  line is required in the spectral fits. The parameters of 
the line obtained from the spectral fit to each observation are reported in Table 
\ref{tab:line}. Note that these are obtained from the fits with the phenomenological
models, but no significant differences are found in the spectra where a {\tt comptt}
model is used.
\begin{table*}[htbp]
\centering
\caption{Parameters of the iron line obtained from the spectral fits to the {\it INTEGRAL} and
{\it RXTE} data. Obs. 1, 2, 3 refer to the average {\it RXTE} spectra, while Obs. 3 low and high refer 
to the sub interval presented in Section 3.2.3. Errors are given at the 90\% confidence. }
\begin{tabular}{ccccc}
\hline
\hline
Obs. & E$_{\mathrm{centroid}}$ & Width ($\sigma$) & Flux & Eq. width \\
     &     (keV)               & (eV) &  ($\times 10^{-4}$~ ph/cm$^2$/s) & (eV) \\
\hline
Faint State &     6.5$_{-0.8}^{+0.4}$ &  1683$_{-445}^{+587}$ &   $125_{-39}^{+81}$  & $1560$    \\
Bright State &    7.2$_{-1.3}^{+0.4}$ &  785$_{-691}^{+1299}$ & 50$_{-21}^{+112}$&  535    \\
Ultra Bright State &  6.6$\pm0.2$  & $< 475$    &   43$_{-17}^{+19}$  & 410     \\
Obs. 1 &       6.53$\pm0.12$            & $<518$     &  3.2$_{-0.9}^{+1.1}$     & 385     \\
Obs. 2 &       6.56$\pm0.10$  &  375$_{-363}^{+211}$  & 3.5$_{-0.8}^{+0.9}$ & 469     \\
Obs. 3 &       6.37$_{-0.17}^{+0.13}$  &  419$_{-181}^{+180}$ &    3.5$_{-0.9}^{+1.6}$ & 258     \\
Obs.3 low &    6.44$_{-0.15}^{+0.12}$           & $<641$   &  2.1$_{-0.6}^{+0.8}$  &  388    \\
Obs.3 high &   6.36$_{-0.08}^{+0.13}$  &  460$_{-296}^{+276}$ & 4.7$_{-1.4}^{+2.2}$  &  213    \\
\hline
\end{tabular}
\label{tab:line}
\end{table*} 
One could wonder whether the line is intrinsic to \igr itself, or whether it could 
originate from the Galactic background. The main argument that points towards an 
origin intrinsic to the system is that if the line was due to the Galactic ridge, 
we would expect its flux to be roughly constant. This is obviously not the case
here.  \\
\indent It is interesting to note that in almost all cases, (except in the ``Bright" and "Faint" 
states), the parameters inferred for the line could be indicative of a narrow line, rather 
than a broad line. In fact for both instruments the upper limit on the line width 
indicates that we are limited by the instrumental spectral resolution. The case of the 
faint and bright states seem different since our fits indicate a broad line (Table \ref{tab:line}). 
Our spectral fits to the {\it INTEGRAL} data (Sec. \ref{sec:integspec}) indicate that 
the ``Bright'' state is spectrally intermediate between the ``Faint'' state and the 
``Ultra-bright'' one, as we will discuss further below. In particular in the 
soft X-rays (4--8 keV), a black body component could be present in the spectra of 
the ``Bright'' state, and represents the data well for the faint state. 
In both cases, a fit to the data with a black body and  a Gaussian (besides 
the power law) does not converge on sensitive parameters for either of the components. 
The broad line we found instead could be indicative of a ``mixture'' of  faint black 
body emission (poorly constrained given the 4 keV lower boundary of our fits) and a 
Gaussian line. This possibility is  compatible with the evolution between the three 
{\it INTEGRAL} ``states'', as clearly seen of Fig. \ref{fig:integspec}, where black 
body emission dominates the soft X-ray in the ``Faint state'' (when either no line is needed 
or a very broad one), to the ``Ultra Bright'' state, where 
no black body is detected, and with a good constraint on the line.\\
\section{Discussion}

\subsection{A neutron star primary?}
\begin{figure}[htbp]
\epsfig{file=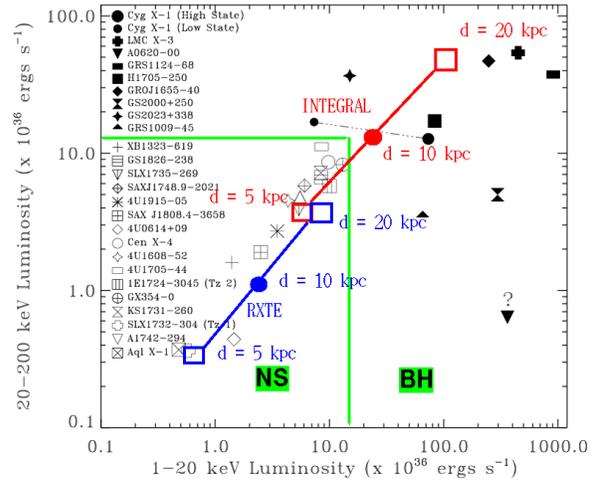,width=\columnwidth}
\caption{\igr luminosities as obtained from the spectral fits presented in this
paper,  and comparison with the ``classification'' proposed in Barret et al. (1996). The two 
continuous lines ending with squares indicate the positions of \igr as obtained in 
this study with {\it INTEGRAL}  and {\it RXTE}  
 assuming three different distances (open squares are at 5 and 20~kpc, filled circles 
are at 10~kpc) for IGR~J19140+0951.
Based on Barret et al. (1996), systems with neutron star primaries would rather lie in the bottom left corner 
(``X-ray burster box'') of the plot, whereas system with black hole binaries would lie outside 
this box (see however di Salvo et al. 2001).}
\label{fig:barret}
\end{figure}

We performed a thorough spectral analysis of the {\it INTEGRAL} source \igr using
 a well-sampled high energy monitoring with {\it INTEGRAL} in 2003 March--May, and 
adding 3 {\it RXTE} observations performed at different epochs. As already observed 
(Paper 1), \igr is highly variable on timescales from months down to hours, and it 
can show variations on shorter timescales as seen during {\it RXTE} observation 3 
(Fig. \ref{fig:PCAHXT}). This behaviour is reminiscent of Galactic X-ray binaries 
(XRB), and our deep analysis further confirms the Galactic nature of IGR~J19140+0951, 
already proposed in other publications (Paper1, Corbet et al. 2004).\\
\indent When observed with {\it RXTE}, the source was dim, with a 1-200 keV (unabsorbed) 
luminosity of $\sim 3.4 \times 10^{36}\times$(D/10~kpc)$^2$ erg/s (Obs.3), and a spectrum 
typical of Comptonisation of soft photons by a low temperature plasma ($kT\sim 5$ keV)
 with a relatively high optical depth ($\tau \sim 5$). This could correspond to the 
``ultra faint state'' which seems to be the state in which the source spends most of 
its time as indicated by our {\it INTEGRAL} monitoring.  During the {\it INTEGRAL} 
observations, the luminosity is up to  about  10 times higher, with a maximum of  
$\sim 3.7 \times 10^{37}\times$(D/10~kpc)$^2$ erg/s. Here significant spectral evolution 
is observed since in one case a bright thermal component may be present in the soft X-rays 
while it is either marginal or  not detected in the other ``states'' defined from the ISGRI light curve. A 
clear pivoting between the three {\it INTEGRAL} spectra is clearly visible (Fig. 
\ref{fig:integspec}). 
The phenomenological models may  indicate a spectral transition 
from something resembling a standard soft state to a hard state (Tanaka \& Shibazaki 1996), as seen in
BHC, but the temperature of the black body, and the parameters of the Comptonisation, especially
during the faint {\it RXTE} observations are more comparable to those of a neutron star primary. 
 The spectral parameters obtained 
from our fits during the ``Bright state'' indicate that it is spectrally intermediate 
between the ``Faint'' state and the ``Ultra bright'' state (Tables \ref{tab:faintres} 
and \ref{tab:brightres}). A hint for a black body component is indeed found here, although 
the best fit involves an iron line. The large width of the latter, and the inability 
of our fits to converge when trying to model the spectra with both the line and the 
black body, tend to indicate that the huge line is in fact a mixture of a narrower 
feature with a fainter and cooler thermal component, for which no constraints can be 
obtained with the 4 keV lower boundary of our spectral analysis.\\
\indent In order to try to constrain the primary type, we first compare the source 
luminosity with that of other known Galactic XRBs. To do so we plotted the 20-200 keV 
vs 1-20 keV luminosities for the brightest {\it INTEGRAL} and {\it RXTE} states, 
at three different distances and over-plotted it with those presented in Barret et al. 
(1996) (Fig. \ref{fig:barret}).

It is clear from Fig. \ref{fig:barret} that, unless the source is at a large distance 
of 20 kpc or beyond, it always falls in the ``X-ray burster box'', except for the 
{\it INTEGRAL} point at 10 kpc, which is exactly half 
way between the 2 standard states of Cyg X-1. However the delimitation between the 
two regions is purely empirical
and based on measurements made up to 1996, on a sample of X-ray bursters only for the 
neutron stars (which at the 
time of  writing were the only known neutron star X-ray binaries with hard X-ray tails 
extending to at least 100 
keV). Since then, Di Salvo et al. (2001) have indeed shown that some neutron star systems, 
could definitely lie outside this so called 
``X-ray burster box''. Therefore unless a very high luminosity hard tail is found, 
the fact that a source lies 
outside the ``X-ray burster box'' is not a definite proof for a black hole binary (Di Salvo et al. 2001).
 In addition, the spectral parameters we obtain from our spectral  fits in all ``states''
 are radically different
from those usually observed in black hole binaries (e.g. McClintock \& Remillard 2004), 
even in their quiescent states 
(e.g. Kong et al. 2002). This is particularly true for the parameters of the cut-off 
energy, or equivalently the electron 
temperature which are in agreement with those presented by Barret (2000) in the case of 
neutron star primaries.\\
\indent As already pointed out in another system (4U 2206+54 Torrej\'on et al. 2004), 
we note that during the ``Faint'' 
state the black body temperature is high, while the source luminosity is not very high 
(although higher than in 
4U 2206+054). Following the procedure presented by Torrej\'on et al. (2004), we can estimate 
the radius of the black body emitter following
 $R_{bb}=3\times 10^4 \times{\mathrm{d_{kpc}}}\sqrt{f^{\mathrm{bol}}/(1+y)}/{\mathrm{kT_{bb}}}$
(in't Zand et al. 1999), where y is the Compton parameter $y\propto\mathrm{kT_{e}\tau^2}$, 
$f^{bol}$ the ``bolometric'' flux
and kT$_{\mathrm{bb}}$ the black body temperature. Using the values found in our study 
(expanding the flux to the 
0.1-200 keV range following in't Zand et al. 1999), we obtain R$_{bb}$=0.999$\times 
\mathrm{d_{kpc}}$ km. We remark here 
a factor of 2 discrepancy between Torrej\'on et al. (2004) and in't Zand et al. (1999), the values 
given in the former are the diameter of the black body emitter, but this does not change 
their conclusions. This value implies that even at a very far distance (e.g. 20 kpc, 
therefore outside of our Galaxy, which appears 
rather unlikely), the black body radius is consistent with the radius of a neutron star.  

\subsection{Possible type of the system}
The absorption column density ($N_{\mathrm H}$=$\sim3$--$10 \times 10^{22}$ cm$^{-2}$) of \igr derived from the 
spectral fits to the {\it RXTE} data is much higher than the Galactic absorption towards the source 
 (1.26$\times10^{22}$ cm$^{-2}$ Dickey \& Lockman, 1990). This favours an absorption intrinsic to the object,
and therefore the presence of absorbing material in the vicinity of the compact object. The variations
of the absorption (Fig. \ref{fig:contour} and Table \ref{tab:RXTE}) also point toward an absorption intrinsic 
to the source. This is in fact similar to what is observed in IGR J16320-4751 (Rodriguez et al. 2003b), or 
4U 1700$-$37 (Boroson et al. 2003),  in which the absorption 
is seen to vary by a factor of about 2 in the former  source (a most likely High Mass X-ray Binary HMXB; 
Rodriguez et al. 2003b) 
and 10 in the latter (a dynamically confirmed HMXB).   The presence of absorbing material is 
consistent with the detection of a (cold) iron line. It is interesting to note that although 
when comparing Obs. 1, 2 and 3, the iron line fluxes are all comparable within the uncertainties 
(Table \ref{tab:line}), while $N_{\mathrm H}$ varies significantly, the line flux is much stronger 
in Obs. 3 high (interval 2), than in Obs. 3 low (interval 1), i.e. it is stronger here when 
$N_{\mathrm H}$ is higher. In addition, there seems to be a tight correlation between the 
1-200 keV (unabs.) flux of \igr and the flux of the line although the case of the 
``Faint'' state does not obey this relation, and the parameters of the line are poorly 
constrained in the ``Bright'' state. This relation, and the relative constancy of the line
energy in most of the cases suggest that the line is produced through fluorescence in a cold medium
as in e.g. Vela X-1 (Ohashi et al. 1984). In addition, the intensity of the iron line during 
the {\it INTEGRAL} observations is comparable to the intensity observed in the HMXB 
GX 301$-$2 at a similar flux (Saraswat et al. 1996).
In the latter system the line width (measured with {\it ASCA}) was consistent with the instrumental 
spectral resolution, which seems to be the case in IGR~J19140+0951, although the energy resolution of both {\it RXTE}/PCA and
{\it INTEGRAL}/JEM-X is very poor in comparison to that of {\it ASCA}/SIS. These similarities between different 
systems would tend to indicate \igr is an HMXB, rather than a system containing a low-mass secondary star (LMXB). 
Finally we observe that the hardest spectra (i.e. those for which 
the electron temperature 
or the cut-off energy is the highest) are observed at higher luminosities, which again is very similar to 
the HMXB 4U 2206+54 (Masetti et al. 2004), 
and rather contrary to what observed in the case of LMXB (Barret 2001).\\
\indent Independently, the temporal variability on timescale $\sim$1000~s is very similar to the HMXBs
 4U 2206+054 (Nereguela \& Reig 2001),
2S 0114+65 (Yamauchi et al. 1990), and Vela X-1 (Kreykenbohm et al. 1999). In these systems, this variability is 
commonly interpreted as due 
to random inhomogeneities in the accretion flow (e.g. Masetti et al. 2004 and references therein). 
The level of variability 
from 0.06 Hz on is compatible with what was found in 4U 1700-37 (Boroson et al. 2003) or 4U 2206+54 
(Nereguela \& Reig 2001), i.e. 
the variability is compatible with purely Poisson noise. In the former source significant aperiodic
variability is detected only below 0.01~Hz, although a 13~mHz QPO is detected at a fractional amplitude 
4.0~\% (Boroson et al. 2003). 
As discussed in Sec. 3.1, if such a feature was present in
IGR~J19140+0951, it should have been 
detected at 
least in {\it RXTE} Obs. 3. 
In 4U 2206+54, on the other hand, significant aperiodic variability is seen below $\sim 0.06$~Hz. However, no 
QPOs are detected in this system. 
Again the similarity of the behaviour of \igr with that of confirmed HMXB, would tend to argue in favour of a 
high mass secondary star in \igr and therefore X-ray luminosity due to wind accretion onto the compact object.\\
\indent The hypothesis of \igr being a HMXB is again in good agreement with the relatively large value of the 
orbital 
period of 13.55 days (Corbet et al. 2004), since HMXBs have usually higher orbital period than LMXBs. Note that  
this is not a 
definite proof since some LMXB can have large orbital period as e.g.  GRS 1915+105 and GRO J1744-28 with  
$\sim 33$ days and  $\sim 12$ days,  
 respectively. The fact that the modulation is sinusoidal (Corbet et al. 2004) would tend to indicate a 
high inclination system (i.e. the orbital plane almost parallel to the line of sight) 
rather than variations of the X-ray flux due to perigee passage of the
compact object in a highly eccentric
 orbit. \\
\indent Finally, it should be noted that \igr lies in the direction of  the Sagittarius arm of our Galaxy, 
which is a region rich of high mass/young stars, and therefore HMXBs. This location could provide 
another indirect support for \igr being a  HMXB, as proposed for 3
similar sources lying in the Norma arm (Revnivtsev 
2003).  This arm is located about 2 kpc from the Sun, and if \igr was  associated with 
 this region its luminosity as obtained from our spectral fit (10$^{35}$-10$^{36}$ erg s$^{-1}$) 
would be completely consistent with that of the aforementioned HMXB/neutron star binaries, as Vela X-1.

\section{Conclusion}
We have presented a detailed study of the hard X-ray properties of \igr observed at different
 times with {\it INTEGRAL} and
{\it RXTE}. From a  well-sampled monitoring of the source in 2003 March--May, we deduced that \igr  
spends most of its time
in a low luminosity state, which likely corresponds to the state observed with {\it RXTE} on three occasions.
The source spectrum is characteristic for thermal Comptonisation, and on one occasion we have evidence 
for a black body 
component in the spectrum. From the comparison of the spectral properties of \igr with those of other XRBs, we 
suggest that this system hosts a neutron star rather than a black hole.\\
\indent The source spectra show evidence for a variable intrinsic absorption which indicate that the compact source
is embedded in a dense cloud. This and the detection  in all our spectra of a
 bright (and thin) iron line, whose flux is higher in the higher luminosity states points towards 
 radial accretion from a stellar wind. 
Therefore it
is very likely that \igr is a HMXB, with properties similar to those of other well known 
HMXB.\\
\indent The arguments presented in the present study are, however, only indicative, none of 
them being definite. 
In particular the identification of counterparts at other wavelengths of the electromagnetic 
spectrum should allow one to truly confirm the nature of the system and/or the compact object. 
Such a study is, however, not possible at the moment given 
the relatively large error on the position of the source. Observations with high resolution
 X-ray satellites, such as {\it Chandra} or {\it XMM-Newton}, should permit a better position to be found, 
counterparts to be searched for, and possibly determine whether \igr is indeed the same 
source as EXO~1912+097. 
In addition such a study should permit one to obtain much better constraints on the absorption and line parameters.

\begin{acknowledgements}
The authors acknowledge M. Cadolle-Bel, A. Goldwurm,  A. Gros, P. Kretschmar, A. Paizis, 
S. Mart\'{\i}nez N\'u\~nez \& R. Walter
for  useful discussions and help with the INTEGRAL data reduction, and
the anonymous referee for helpful comments, and the suggestion of
describing the 
spectral extraction in more detail, which helped to improve the paper. We also acknowledge 
P. Lubinski for useful discussions on the spectral extraction methods, and for 
sharing some results with us prior to publication. JR is 
very grateful to T.E. Strohmayer and the RXTE help desk for appreciable help
on the PCA confusion issue, and J. Swank for triggering the RXTE 2003 ToO. 
JR acknowledges financial support from the French space agency (CNES). 
DCH gratefully acknowledges a fellowship from the Academy of Finland. SES
is  supported by the UK PPARC. 
JS acknowledges the financial support of Vilho, Yrj\"o and
Kalle V\"ais\"al\"a foundation and is grateful to  
the Finnish space research programme Antares and TEKES. This paper
is based on observations with INTEGRAL, an ESA project with instruments
 and science data centre funded by ESA member states (especially the PI 
countries: Denmark, France, Germany, Italy, Switzerland, Spain), Czech 
Republic and Poland, and with the participation of Russia and the USA. 
This research has also made use of data obtained through the High Energy 
Astrophysics Science Archive Center Online Service, provided by the NASA 
Goddard Space Flight Center
\end{acknowledgements}

\section*{Appendix 1: ISGRI Spectral extraction from images}
\subsection*{Principle of the method}
Numerous issues with OSA, which remains under development, are
       reported on the ISDC website'\footnote{See the known issues at:\\ 
http://isdc.unige.ch/Soft/download/osa/osa\_sw/osa\_sw-4.1/osa\_issues.txt.}
       Of particular relevance to this work : 
       ``ii\_spectra\_extract runs per
 science window and in case of weak
       sources, addition of many spectra obtained for the different
       science windows may give a bad total spectrum. Spectral
       reconstruction is very sensitive to the background correction.
       In certain cases running the imaging procedure on several
       (large) energy bands can provide a better spectrum'' .  \\
We therefore extracted the spectra of IGR J19140+0951 using a method based 
on the count rates extracted from the images. 
For the whole data set (only restricted to the SCW where the source is 
less than 5$^\circ$ from the center of the FOV), we ran the software up to the IMA level.
We extracted the products  20 energy bins defined such
that they match exactly the boundaries of the redistribution matrix file (rmf).
The energy bands are 20.65-24.48, 24.48-28.31, 28.31-32.14, 32.14-35.97, 35.97-39.8, 39.8-43.36,
 43.36-49.38, 49.38-53.21, 53.21-57.04, 57.04-60.87, 60.87-68.52, 68.52-76.18, 76.18-87.67, 
87.67-99.16, 99.16-122.14, 122.14-150.86, 150.86-196.82, 196.82-300.22, 300.22-518.5, 
518.5-1000 keV. 
Note that the energy ranges above $\sim 300$ keV are of limited use for most of the sources.\\
\indent Once this is finished, for each SCW,  
 the intensity, exposure, variance and significance 
maps of the field are obtained in each of the aforementioned energy ranges.
The average count rate, $F(E_i,\alpha,\delta)$, in the energy range $E_i$, over a list 
of $p$ SCW, at the position ($\alpha$, $\delta$) of a given source is given by:
\begin{equation}
F(E_i,\alpha,\delta)=\frac{\sum_{j=1}^{p}\frac{F_j(E_i,\alpha,\delta)}{var_j(E_i,\alpha,\delta)}}{\sum_{j=1}^{p}\frac{1}{{var_j(E_i,\alpha,\delta)}}}
\label{eq:mosa}
\end{equation}
where $F_j(E_i,\alpha,\delta)$ is the count rate in SCW \#j, in the energy range $E_i$ at a (sky) position
($\alpha$, $\delta$), and $var_j(E_i,\alpha,\delta)$ is the associated variance value.\\
Repeating Eq. \ref{eq:mosa} from i=1, to i=20 (in our case) allowed us to obtain the 
source spectrum over the given list of SCW.
\subsection*{Validity of the method: estimate of a Crab spectrum}
In order to validate our method, we extracted a Crab spectrum following the same method.
In order to be even more rigourous, we restricted our comparison to Crab observations
performed with {\it the same observing pattern as most of our observations}, i.e.
a hexagonal pattern. However we point out that a check on an arbitrary pattern gave similar
and consistent results. The validation of the spectral extraction method is currently a work in progress
and detailed results and issues will be presented in a forthcoming paper (Lubinski et 
al. in prep.). In general, and for what concerns this work,  the discrepancy 
between the standard spectral extraction and this new method does not exceed 5\% 
(Lubinski private comm.). Our particular spectral analysis of the Crab using both methods
showed that the spectral parameters were compatible within  1\% for the photon index,
within about 5 \% for the normalization, and the 20-200 keV flux discrepancy 
is about 2\%.  
\end{document}